\documentclass[english,english,11pt]{article}
\usepackage{amstext,amsmath,amssymb}
\usepackage[OT1]{fontenc}
\usepackage[latin9]{inputenc}
\usepackage[english]{babel}
\usepackage{float}
\usepackage{graphicx}
\usepackage{geometry}
\usepackage{ulem}
\usepackage[comma,square,numbers,sort&compress]{natbib}
\usepackage[labelfont=bf,font=small]{caption}
\usepackage[displaymath,mathlines]{lineno}

\usepackage[usenames,dvipsnames]{color}

\newcommand{\mroy}[1]{\textcolor{black}{$^{\textrm{}}${#1}}}
\newcommand{\mren}[1]{\textcolor{black}{$^{\textrm{}}${#1}}}
\newcommand{\jt}[1]{\textcolor{black}{$^{\textrm{}}${#1}}}
\newcommand{\mre}[1]{\textcolor{black}{$^{\textrm{}}${#1}}}
\newcommand{\mroys}[1]{\textcolor{black}{$^{\textrm{}}${#1}}}
\newcommand{\roy}[1]{\textcolor{black}{$^{\textrm{}}${#1}}}
\newcommand{\mrsept}[1]{\textcolor{black}{$^{\textrm{}}${#1}}}
\newcommand{\mroysept}[1]{\textcolor{black}{$^{\textrm{}}${#1}}}
\newcommand{\roysept}[1]{\textcolor{black}{$^{\textrm{}}${#1}}}

\usepackage{fixltx2e}

\begin{document}

\title{Methane mitigation timelines to inform energy technology evaluation}

\author{Mandira Roy\thanks{Engineering Systems Division, Massachusetts Institute of Technology,  77 Massachusetts Avenue, Cambridge, MA02139, USA.}, Morgan R. Edwards$^*,$ Jessika E. Trancik$^*$\thanks{Santa Fe Institute, 1399 Hyde Park Road, Santa Fe, NM 87501, USA.} \thanks{Email: trancik@mit.edu}}



\maketitle


\begin{abstract}
Energy technologies emitting differing proportions of methane (CH$_4$) and carbon dioxide (CO$_2$) vary \jt{significantly} in their relative climate impacts over time, due to the distinct atmospheric lifetimes \jt{and radiative efficiencies} of the two gases. Standard technology comparisons using the global warming potential (GWP) \jt{with a fixed time horizon do not account for the timing of emissions in relation to climate policy goals.} Here we develop a portfolio optimization model that incorporates changes in technology impacts \mre{based on the} temporal proximity \mre{of emissions} to a radiative forcing (RF) stabilization target. An optimal portfolio, maximizing allowed energy consumption while meeting the RF target, is obtained by year-wise minimization of the marginal RF impact in an intended stabilization year.  The optimal portfolio calls for using certain higher\mrsept{-}CH$_4$-emitting technologies prior to an optimal switching year, followed by CH$_4$-light technologies as the stabilization year approaches. We apply the model to evaluate transportation technology pairs and find that accounting for \jt{dynamic emissions impacts, in place of \mroys{using} the static GWP, can result in CH$_4$ mitigation timelines and technology transitions that allow for significantly greater energy consumption while meeting a climate policy target. The results can inform the forward-looking evaluation of energy technologies by engineers, private investors, and policy makers.}  
\end{abstract}


\section{Introduction}

\jt{Energy technologies emit greenhouse gases, primarily CO$_2$ and CH$_4$, with widely differing atmospheric lifetimes \mroys{and radiative efficiencies \cite{IPCC2013}}. The temporal proximity of these emissions to a climate policy threshold, such as a radiative forcing (RF) stabilization \mroys{target}, should factor into assessments of the climate impacts of energy technologies. This is because\mre{,} in the presence of an RF stabilization policy \cite{UNFCCC1992,UNFCCC2009}, the importance of mitigating \mroys{more potent but} shorter-lived greenhouse gases will increase as the build-up of forcing agents approaches the target level \mre{\cite{Jackson2009,Smith2012,Shindell2012,Shoemaker2013b}}.} \jt{Standard technology evaluation does not account for the timing of emissions relative to a climate policy goal. The most commonly used \mre{method} converts different greenhouse gases to their CO$_2$-equivalent mass values using the GWP(100) emissions equivalency metric \mre{\cite{Rodhe1990,Shindell2009}},
which compares gases by integrating their RF impacts over a fixed time horizon \mre{of 100 years}.} \jt{Despite scientific and economic critiques of the static GWP(100) }
\mre{(e.g. \cite{ONeill2000,ONeill2003,Shine2009,Daniel2011,Peters2011, Stratton2011, Kendall2012})},
this metric is widely applied in forward-looking technology evaluation \mre{\cite{IPCC2012}} and in climate change mitigation policy formulation \mre{\cite{UNFCCC1992,Whitehouse2013,EPA2014a,EPA2014b,EOP2014}}.

\jt{The dynamic climate impacts of technology adoption scenarios can be studied using integrated assessment models \mre{\cite{vanVuuren2006,Weyant2006,vanVuuren2007,Rogelj2011,Smith2013_1,Rogelj2014}}, which capture the interdependencies of gases that are co-emitted by various technologies in differing proportions \mre{\cite{Smith2013_2,Rogelj2014}}. 
\mre{Models} also demonstrate the benefits of mitigating short-lived climate forcings to reduce peak warming, and (in contrast to the benefits of immediate CO$_2$ reductions) find more limited benefit\mre{s} from early mitigation of these forcing agents \mre{\cite{Bowerman2013,Shoemaker2013}}. To be applied to technology evaluation, however, these insights must be translated from the level of the scenario or set of scenarios modeled (measuring total impact) to the level of technologies (measuring impact intensities, e.g. emissions per unit energy converted) 
\cite{Trancik2013,Trancik2014}. Given the advance planning needed for technology development, to support R\&D and infrastructure investment, and the inherent uncertainties about the future scenario to be followed (e.g. energy demand, emissions pathways), these impact intensity estimates must be reasonably robust to a range of possible future scenarios.}

\jt{In recognition of the 
\mrsept{need for} simple tools to perform dynamic comparisons of emissions impacts, several  emissions equivalency metrics have been proposed as dynamic alternatives to the GWP \mre{\cite{Manne2001,Shine2007,IPCC2009,Johansson2012,Tanaka2013,Berntsen2010,Alvarez2012,Edwards2014}}, based on instantaneous and integrated measures of temperature, RF, and economic impact. Questions remain about how beneficial these alternatives are, and whether they can be used to make technology comparisons that are robust to a range of future scenarios. Here we contribute to this debate by formulating a model to investigate optimal technology choice under an RF constraint, \mroys{identifying the corresponding metric}, and showing the benefits of applying this method to technology evaluation. We answer the following questions. Given the emissions intensities of candidate technologies, how much is gained by applying a dynamic metric (as compared to the static GWP)?} 
\roysept{Can a scenario-independent technology comparison be performed?}

We represent dynamic technology choice as a simplified forward-looking multi-period portfolio optimization problem, maximizing energy consumption over a planning horizon in the presence of an RF stabilization constraint. This formulation leads naturally to an analytical expression for technology impact that changes over time based on the marginal RF impact in the stabilization year. \mrsept{The marginal RF impact can be determined for a range of scenarios leading to stabilization at a given target RF level (Sec.\ S1).} 
This formulation of the technology choice problem is equivalent to applying the instantaneous climate impact (ICI) emissions equivalency metric proposed in earlier work \cite{Edwards2014}. \jt{Here we demonstrate} \jt{the sizable benefits of using this approach to plan for technology transitions} \jt{given a global RF target. The resulting technology evaluation is robust to uncertainties in the stabilization year \mrsept{over a range of scenarios}, as well as to uncertainties in future radiative efficiencies and atmospheric lifetimes of CH$_4$, under a 3 W/m$^2$ RF target.}

The optimal technology portfolio uses relatively CH$_4$-heavy technologies in earlier years, switching to relatively CH$_4$-light technologies as an intended RF stabilization year approaches.  This switching portfolio facilitates significantly greater energy consumption than the exclusive use of either technology alone. These results suggest a role for CH$_4$-heavy technologies as ``bridges'' to lower emissions intensity alternatives. The early use of the CH$_4$-heavy technology (the first listed in each pair) is optimal only if the stabilization horizon exceeds 22 years for compressed natural gas and gasoline, 14 years for algae biodiesel and electric vehicles, and 19 years for renewable natural gas and switchgrass ethanol. Given a stabilization horizon from the present to mid-century, the \mren{energy} consumption gain from an optimal switching portfolio can be up to 15$\%$ and 50$\%$ compared to using only a CH$_4$-light or CH$_4$-heavy technology, respectively. 
The GWP\mren{($\tau$), in contrast, leads to a single, static technology portfolio, which for the GWP(100) results}
in either a significant overshoot of the stabilization target or, if constrained by the target \roysept{(for example through a multi-basket emissions policy that addresses greenhouse gases separately)} allows significantly lower energy consumption. \jt{The GWP(35) does not lead to an overshoot but \mrsept{results in} 
lower energy consumption than the switching portfolio.} 

\jt{The main contributions of this paper are twofold. We show that the differences in the CH$_4$ and CO$_2$ emissions intensities of the transportation technologies examined are large enough that planning for technology transitions and CH$_4$ mitigation can yield significant returns, in terms of supporting energy consumption while meeting a climate policy target (here formulated around \mren{RF} stabilization). We also develop a method for technology evaluation against climate policy targets that allows for an effective and relatively scenario-independent technology comparison, as described further in sections to follow.}

\section{Methods}\label{methods}

In this section we describe the sectoral RF stabilization target (Sec.\ \ref{horizon}), the evaluation of technology RF impacts (Sec.\ \ref{emissions}-\ref{technologies}), and the technology portfolio optimization model (Sec.\ \ref{certainty}).  
    
\subsection{RF stabilization constraints} \label{horizon}   

\mroys{Changing atmospheric concentrations of greenhouse gases are associated with changes in global mean temperature (with a time lag) and a range of impacts related to temperature or more directly to heat fluxes \cite{Lenton2011}. 
Climate mitigation targets are commonly formulated around a recommended temperature threshold~\cite{UNFCCC2009}, from which an equilibrium RF stabilization level can be derived.}  We use a 3 W/m$^2$ global RF stabilization target, which in equilibrium is roughly equivalent to a 2$^{\circ}$C temperature change~\cite{IPCC2013} from pre-industrial levels, a commonly-cited climate target~\cite{UNFCCC2009}. A range of scenarios stabilizing at this level is  determined~\cite{Allen2009,Edwards2014}, with stabilization occurring within a range of approximately 15 years up to \mren{2050} (see Sec.\ S1). A stabilization year of 2050, consistent with a 3 W/m$^2$ RF target and the RCP\mren{2.6} 
scenario~\cite{Meinhausen2011,vanVuuren2011b}, is used as an example, but we also examine the effect of earlier stabilization years. 

Beginning in the year 2015, the global RF target required for stabilization in 2050, computed by subtracting the estimated RF due to legacy emissions (pre-2015 emissions remaining in the atmosphere in 2050) from 3 W/m$^2$, is found to be 1.6 W/m$^2$ \mrsept{of which an estimated 70\% or 1.12 W/m$^2$ is attributable to global energy\mroysept{-}related emissions \cite{IPCC2013}}. In our model, the RF stabilization target for a specific energy sector is its fraction of the global \mrsept{energy\mroysept{-}related RF} target in proportion to its energy consumption today relative to total global energy consumption. The US road transportation sector constitutes about 4$\%$ of today's global primary energy consumption \cite{EIA2014}. We consider
\mrsept{as an example, 36$\%$} of the US road transportation sector for our technology portfolio choice and an RF stabilization target (TRF) for this subsector of 1$\%$ of the global stabilization target, or 0.016 W/m$^2$. The benefits of planning for CH$_4$ mitigation would apply to larger energy end-use sectors as well. (\jt{The effect on our results of deviating from this particular sectoral RF target is discussed in Sec.\ \ref{results2}} \mroys{under \text{\mre{Sensitivity to} sectoral RF target}}.)

\subsection{Marginal RF and GWP calculations} \label{emissions}

Technologies emit multiple greenhouse gases, the three most significant being CO$_2$, CH$_4$ and nitrous oxide (N$_2$O), indexed by $i=K, M, N$, respectively.  The RF following the use of a technology can be linearly approximated by a function of the emission intensities of these gases, and their radiative efficiencies and atmospheric lifetimes \cite{IPCC2013}. \mroys{However, in Secs.\ \ref{results2} and S10 the effects of variable gas lifetimes and non-linearities in this relationship, due to the effect of changing atmospheric greenhouse gas concentrations on marginal RF, are studied \cite{IPCC2013Ch8}.} Let $b_{ij}$ denote the mass of gas $i$ emitted by technology $j$ per unit \mren{energy} consumption, $A_i$ the radiative efficiency of gas $i$, and $f_i(t,t')$ the impulse response function representing the fraction of gas $i$ retained in the atmosphere at time $t$ following emission at time $t'$,
  
\begin{equation}
f_i(t,t')=\text{exp}\left(-\frac{t-t'}{\tau_{i}}\right), \ \text{for} \ i=M,N,
\end{equation}
\begin{equation}
f_K(t,t')=a_0 + \sum_{k=1}^{3}a_{k} \cdot \text{exp}\left(-\frac{t-t'}{\tau_{k}}\right).
\end{equation}
Empirical values of $A_i$ and the parameters in $f_i(t,t')$ for CO$_2$, CH$_4$ and N$_2$O \cite{IPCC2013} are given in Sec.\ S2. 

The instantaneous RF from unit \mren{energy} consumption using technology $j$ is $\sum_{i}b_{ij}A_i$ 
and the RF impact at evaluation time $t$ of a pulse emission from unit \mren{energy} consumption using technology $j$ at emission time $t'$ is
  
\begin{equation}\label{rfj} 
RF_j(t,t')= \sum_i b_{ij} A_i f_i(t,t').
\end{equation}

For sustained emissions occurring over time, prior to the evaluation time $t$, $RF_j(t,t')$ in \eqref{rfj} represents the marginal RF impact at $t$ of unit \mren{energy} consumption at emission time $t'$. This corresponds to the absolute ICI metric~\cite{Edwards2014} for technology $j$ 
(see Sec.\ S5).  

Using the same parameters, technology $j$'s impact using the GWP($\tau$), is

\begin{equation}
GWP_j(\tau)=\sum_i b_{ij} GWP_i(\tau),\label{gwpj}
\end{equation}   

\noindent in grams CO$_2$-equivalent per unit energy consumption, where

\begin{equation}
\label{gwpghg}
GWP_i(\tau)=\frac{\int_{t'}^{t'+\tau} A_i f_i(t'',t')dt''}{\int_{t'}^{t'+\tau} A_K f_K(t'',t')dt''}
\end{equation}

\noindent 
and $\tau$ is the integration horizon.

\subsection{Description of technologies}\label{technologies}

Life cycle emissions intensities $b_{ij}$ are obtained from the
GREET model (https://greet.es.anl.gov) for three pairs of transportation technologies, where `technology' refers to the combined fuel and vehicle jointly. \mroy{Emissions intensities are current estimates for the US, and values may vary geographically and over time.  
Significant reductions in CH$_4$ emissions may be possible \cite{EPA2013}. The sensitivity of our results to such variations is discussed in Sec.\ \ref{results2}} \mroys{under \text{Effect of variable emissions intensities}.}

\begin{figure}[!htp]
\centering
\includegraphics[width=110mm]{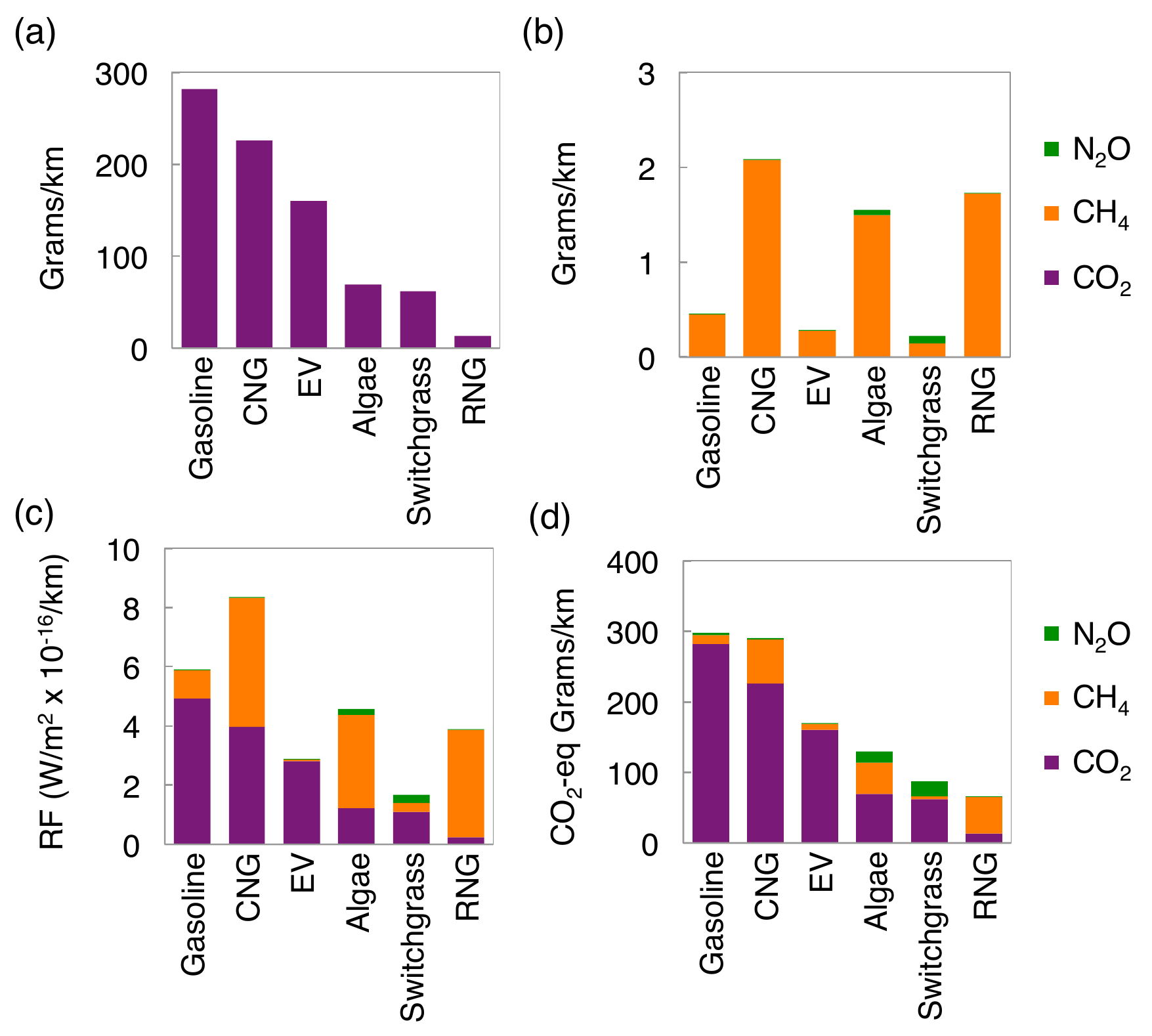}
\caption{\small{Life cycle technology greenhouse gas emissions per kilometer traveled. (a) Grams CO$_2$ emitted. (b) Grams CH$_4$ and N$_2$O emitted. (c) Instantaneous radiative forcing. 
(d) Grams CO$_2$-equivalent per kilometer using GWP(100) for each technology.  Technologies vary in the composition of their emissions.  Gasoline, electric vehicles (EV, using current U.S. electricity mix), and an 85\% switchgrass ethanol blend with gasoline are CH$_4$-light and compressed natural gas (CNG), algae biodiesel, and landfill renewable natural gas (RNG), all with CH$_4$ leakage, are CH$_4$-heavy. (Data source: GREET 2013.)}} 
\label{fig:irf}
\end{figure}

For each technology pair the CH$_4$-heavy $h$ and CH$_4$-light $l$ are chosen so that $b_{Mh}>b_{Ml}$ (Fig.\ \ref{fig:irf}(a)-(b)). CH$_4$-heavy technologies \roy{compressed natural gas} (CNG), \roy{algae biodiesel} (algae), and \roy{renewable natural gas} (RNG) exhibit higher instantaneous RF than their CH$_4$-light counterparts gasoline, electric vehicle (EV), and switchgrass ethanol (switchgrass) (Fig.\ \ref{fig:irf}(c)) but lower GWP(100)-based impacts (Fig.\ \ref{fig:irf}(d)), due to the higher radiative efficiency but faster decay time of CH$_4$ relative to CO$_2$. See Tab.\ S3 for numerical values associated with Fig.\ \ref{fig:irf}(a)-(d).

\begin{figure*}[htbp]
\centering
\includegraphics[width=155mm]{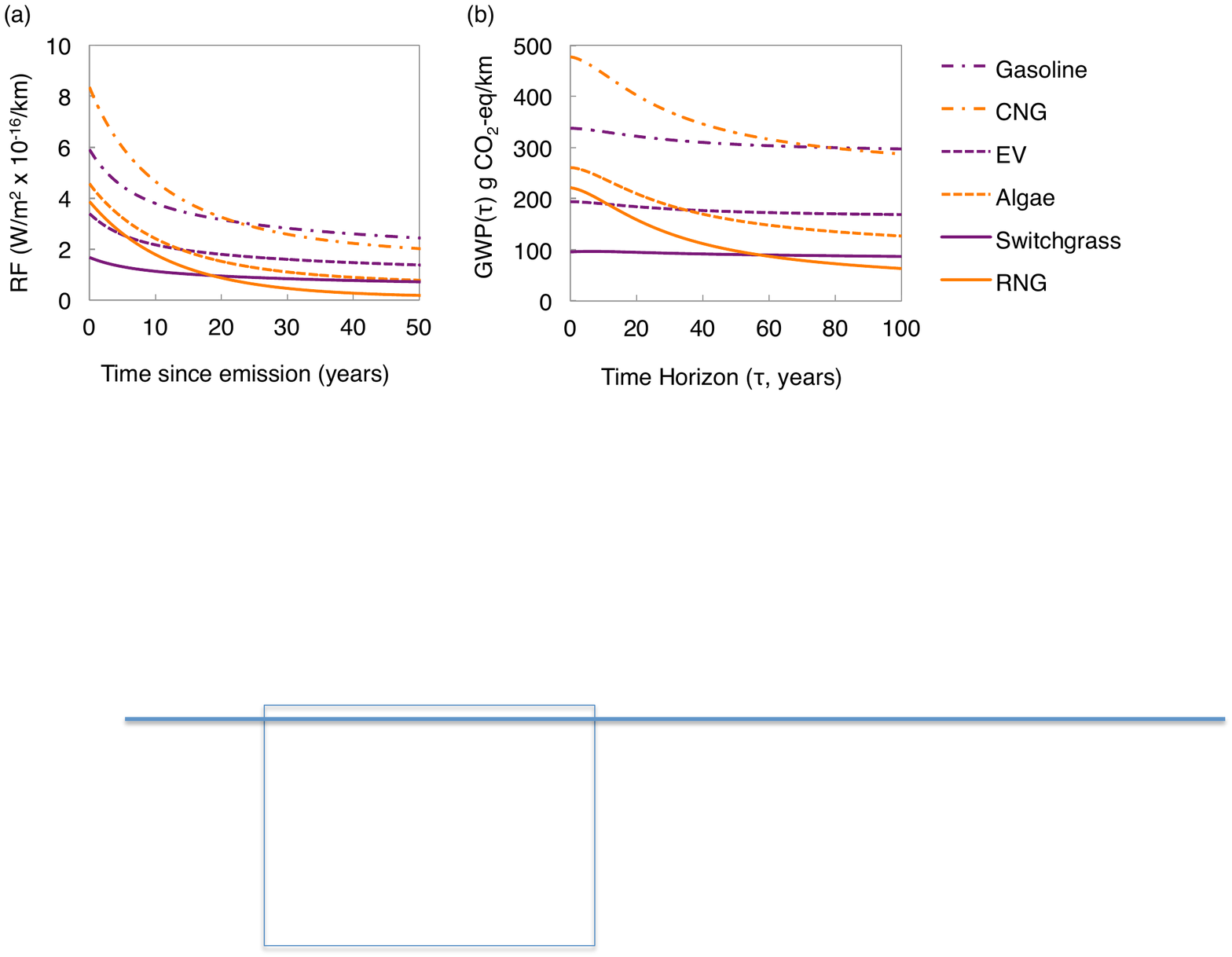}
\caption{\small{\mren{RF} and global warming potential (GWP) of technology emissions. (a) RF over time resulting from use of a technology at $t=0$. (b) GWP over integration horizon $\tau$ resulting from use of a technology at $t=0$. Purple represents CH$_4$-light fuels: gasoline, electric vehicles (EV), and switchgrass.  Orange represents CH$_4$-heavy fuels: compressed natural gas (CNG), algae, and landfill renewable natural gas (RNG). Fuels with similar GWP(35) values are presented in pairs, with solid, dashed, or dot-dashed lines.}}
\label{fig:rfandgwp}
\end{figure*}

Since CH$_4$ decays much faster than CO$_2$ (and N$_2$O), the RF induced by an initial pulse emission from a CH$_4$-heavy technology falls faster with $t$ and its GWP($\tau$) falls faster with $\tau$ than for its CH$_4$-light counterpart (Fig.\ \ref{fig:rfandgwp}(a)-(b)). The initial values in Fig.\ \ref{fig:rfandgwp}(a) correspond to the bars in Fig.\ \ref{fig:irf}(c), and the GWP($\tau$) values at $\tau$=100 in Fig.\ \ref{fig:rfandgwp}(b) correspond to the bars in Fig.\ \ref{fig:irf}(d). To the extent possible, technologies in each pair are matched in terms of their GWP(35) impact values, where 35 years is the stabilization horizon between 2015 and 2050. 
    
\subsection{Technology portfolio optimization model}\label{certainty}

Energy technologies emit sustained streams of greenhouse gases over time rather than as a single pulse.  We use a discrete time approximation of energy consumption where emissions occur as a pulse at the end of each year $t'=0,\ldots,t_S$.

Let $c_{t'}$ denote energy demand in year $t'$ and $x_{jt'}$ the fraction of $c_{t'}$ supplied by technology $j$ in year $t'$. A technology portfolio $p$ is defined by the set $x_{jt'}$ over time $t'=0,\ldots,t_S$, and $RF_p(t_S)$ denotes the total \mren{RF} induced by the portfolio at the end of the stabilization year $t_S$. In the model presented here, the technology planning horizon coincides with the RF stabilization year, given the coincidence of commonly suggested stabilization horizons \cite{Meinhausen2011,vanVuuren2011a} and timelines for technology development and infrastructure planning \cite{Williams2012}. However, the model could be adapted to cases where the planning and stabilization horizons differ (see Sec.\ S6.1). The optimization model that selects a technology portfolio and energy consumption levels to maximize total consumption, while satisfying the RF target in the stabilization year, is given below.  

\vspace{2mm}
\noindent\textbf{\textit{Optimization Model}}
\begin{align}
	  &\underset{c_{t'}, x_{jt'} \forall j,t'}{\text{Max}} \quad \sum^{t_S}_{t'=0}c_{t'}   \nonumber \\
		& \text{s. t.} \ \quad c_{t'}=c_{t'-1}(1+g_{t'}) \ \text{for} \ \ 1\leq t' \leq t_S, \nonumber\\
		& \qquad \quad RF_{p}(t_S)\leq TRF, \nonumber \\
		& \qquad \quad x_{jt'} \in [0,1] \ \text{and} \ \sum_{j}x_{jt'}=1, \ 	
		 \label{portwt}   \nonumber\\
		& \qquad\qquad \ \text{for}\ j=h,l\ \text{and} \ 0\leq t' \leq t_S, \nonumber
\end{align}

\noindent where the objective function represents total \mren{energy} consumption, the first constraint defines the energy consumption profile based on growth rate $g_{t'}$ in year $t'$ and the second constraint ensures that RF does not exceed the target (TRF) in the stabilization year.

The portfolio contribution to RF, given by $RF_p(t_S)$, is 

\begin{equation}\label{RFpS}
RF_p(t_S)=\sum_{t'=0}^{t_S} c_{t'} \sum_{j}x_{jt'} RF_j(t_S,t').
\end{equation}

\noindent $RF_p(t_S)$ is the sum of RF impacts in the stabilization year of all prior portfolio emissions.  $RF_j(t_S,t')$ represents the marginal RF impact of unit \mren{energy} consumption using technology $j$ in emission year $t'$. Using \eqref{RFpS} in the model simplifies its solution as we can first determine the optimal technology choice in each year by minimizing $RF_p(t_S)$ and then maximize \mren{energy} consumption using the optimal portfolio.

Since the model constrains RF only in the stabilization year, TRF overshoots are possible.  An additional set of constraints,
 
\begin{equation}
RF_{p}(t)\leq TRF \ \text{for} \ 0<t<t_S, \label{overshoot}
\end{equation}

\noindent (referred to as overshoot constraints) are also presented to assess the impact of overshoot restrictions on optimal \mren{energy} consumption levels. \roysept{Overshoots could be restricted through policies that separately cap short- and long-lived greenhouse gases~\cite{Smith2012}.} 

The model is applied to the three transportation technology pairs shown in Fig.\ \ref{fig:irf}.  We consider annual energy consumption growth rates of 0$\%$, representing flat consumption. (In Sec.\ S6.2, we also consider how the results change under an energy consumption growth rate of 1.2$\%$ 
\cite{EIA2014}.) Additionally, we consider the effect of uncertainty in the stabilization horizon (see Sec.\ \roy{S9}), based on a 15-year range of stabilization years given a plausible set of emissions scenarios for stabilizing \mren{RF} at 3 W/m$^2$ (see Sec.\ \ref{results2} and Sec.\ S1).

\subsection{Portfolio optimization with the GWP} \label{gwp} 

\mroy{We compare the optimal technology portfolio based on the dynamic emissions impact evaluation, with the portfolio (and energy consumption) that would be obtained using the GWP. This allows us to estimate the gains of this approach over the standard GWP-based method for technology evaluation. CO$_2$-equivalent emissions are determined using the GWP and treated as CO$_2$ when evaluating the RF impact in the stabilization year. (This is similar to the approach outlined above but uses the GWP in place of a marginal RF impact based metric.)} Therefore, the GWP-based estimate of RF impact in the stabilization year of technology $j$ per unit \mren{energy} consumption in year $t'$ is $GWP_j(\tau) A_K f_K(t_S,t')$, where $GWP_j(\tau)$ is given by \eqref{gwpj} \mren{(in units CO$_2$-equivalent per unit energy consumption)}, $A_K$ is the radiative efficiency of CO$_2$, and $f_K(t_S,t')$ is the fraction of CO$_2$ emitted at time $t'$ remaining in the atmosphere at time $t_S$. Using this definition, the  
intended RF in year $t_S$, based on GWP, of using technology portfolio $p$ for the \mren{energy} consumption stream $c_0,\ldots,c_{t_S}$, is

\begin{equation}\label{rfp'}
RF_p^g(t_S)=\sum^{t_S}_{t'=0}c_{t'}\sum_{j} x_{jt'}GWP_j(\tau)A_K f_K(t_S,t').
\end{equation}   

\noindent  \jt{Equation \eqref{rfp'} is used in the optimization model in place of $RF_p(t_S)$  given in \eqref{RFpS} to determine the maximum energy consumption allowed by the GWP-based portfolio. 
We consider both GWP(100) and GWP(35) in our numerical simulations: GWP(100) because it is the most widely used metric and GWP(35) because it is consistent with our planning horizon of 35 years.} 

\section{Results}\label{results}

The solutions to the technology portfolio optimization model are described in Sec.\ \ref{results1} where we investigate the benefits of planning for CH$_4$ mitigation. Sec.\ \ref{gwpresults} describes the optimal solutions calculated using the GWP. Sec.\ \ref{results2} presents a sensitivity analysis. 

\subsection{Optimal portfolio \jt{based on dynamic emissions impact evaluation}} \label{results1}

\begin{figure}[htbp]
\centering
\includegraphics[width=95mm]{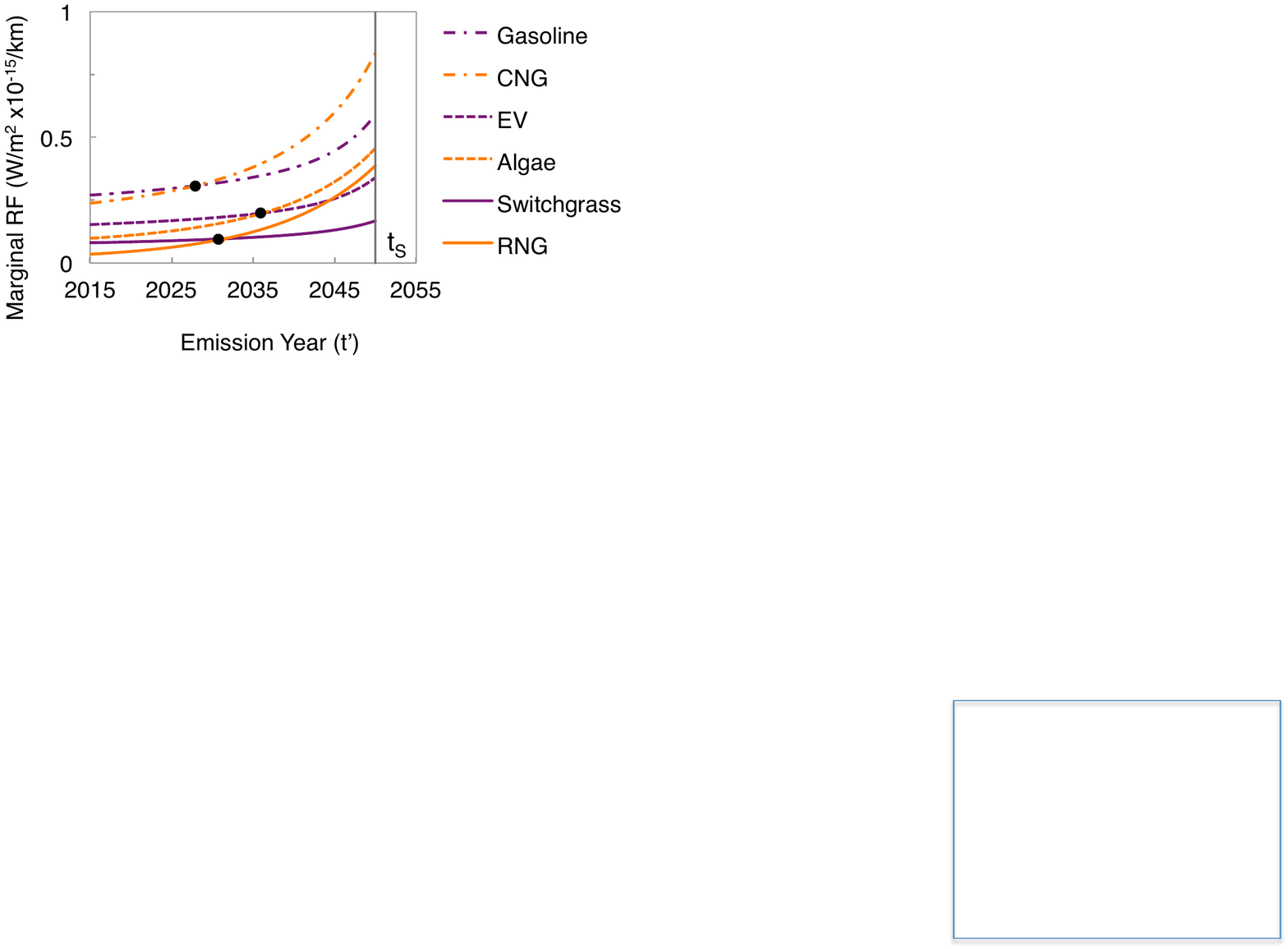}
\caption{\small{Marginal \mren{RF} impact at the time of stabilization (t$_S$) of technology use at different emission times $t'$. Pair-wise comparisons are presented for gasoline and compressed natural gas (CNG), electric vehicle (EV) and algae, and switchgrass and landfill renewable natural gas (RNG), with corresponding switching horizons of 13, 21, and 16 years indicated by black dots.}}
\label{fig:mrfi} 
\end{figure}

The optimal technology portfolio is determined by year-wise minimization of $RF_j(t_S,t')$ across technologies, which can yield a technology portfolio switching from the CH$_4$-heavy to the CH$_4$-light technology in an optimal switching year.  The  
maximum possible energy consumption \mren{is determined} using the optimal technology portfolio. The results support the use of suitable CH$_4$-heavy technologies as bridging technologies, if the stabilization horizon is sufficiently long i.e.,\ $t^*$ exists and has not already passed.  All three technology pairs shown in Fig.\ \ref{fig:irf} satisfy these conditions.  The marginal RF impact resulting from using each technology over time, in an example stabilization year of 2050, is shown in Fig. \ref{fig:mrfi}. The general solution to the optimal portfolio problem is given in Sec\roy{s}.\ S6.1 \roy{and S6.2}.


The optimal switching to stabilization time span is 22 years for CNG and gasoline, 14 years for algae and EV, and 19 years for RNG and switchgrass.  Using a stabilization horizon of 35 years (2015-2050), it is optimal to use CNG for 13 years, algae for 21 years, and RNG for 16 years, followed by a switch to gasoline, EV, and switchgrass, respectively. If the stabilization year is shifted up to 2043 (the middle of the modeled range of years discussed in Sec.\ \ref{horizon}), the optimal switching year for each technology pair shifts by 7 years, keeping the same time span between switching and stabilization \mren{(see Sec. S9)}. 

\begin{figure*}[htbp]
\centering
\includegraphics[width=150mm]{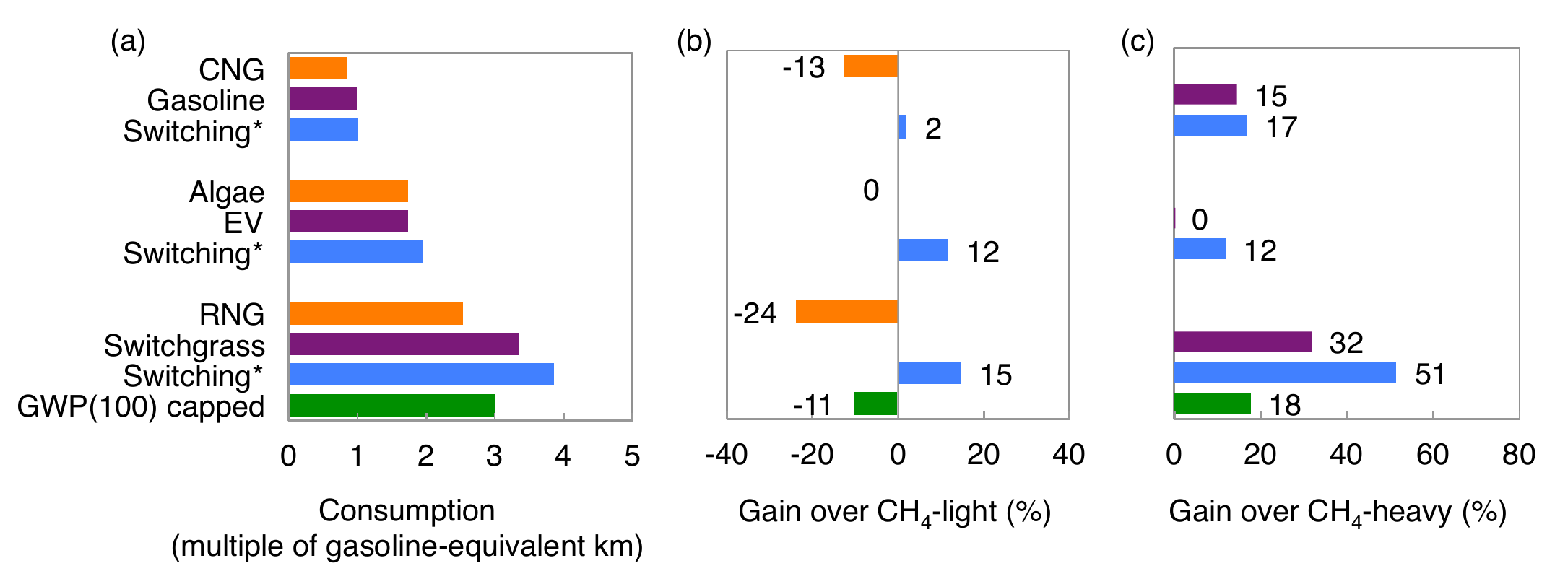}
\caption{\small{Optimal energy consumption over the stabilization horizon that meets the target \mren{RF}  level at the stabilization time. (a) Total consumption for the three technology pairs (as a multiple of travel distance supported by gasoline): 
\mroysept{gasoline and compressed natural gas (CNG), electric vehicles (EV) and algae, 
and switchgrass and landfill renewable natural gas (RNG)}. (b) Percentage gain in energy consumption over the CH$_4$-light (purple) technology. (c) Percentage gain over the CH$_4$-heavy technology (orange). Switching portfolios (blue) involve a transition from a CH$_4$-heavy to a CH$_4$-light technology. The 
GWP(100) capped portfolio (green) selects the CH$_4$-heavy technology \roysept{and approximates a case in which different gases are regulated with separate policies, a `multi-basket' approach, thereby avoiding an overshoot.} \jt{Applying the GWP(35) results in the selection of the CH$_4$-light technology (purple) and corresponding energy consumption level.}
}}
\label{fig:cstar} 
\end{figure*}

We calculate the maximum allowed gasoline-equivalent energy consumption for individual technologies in each technology pair and compare the values to the consumption using the optimal switching portfolio (Fig.\ \ref{fig:cstar}). The optimal switching portfolio  increases the allowed energy consumption relative to each individual technology alone. The percentage \mren{energy} consumption gains relative to the individual CH$_4$-light and CH$_4$-heavy technologies are shown in Fig.\  \ref{fig:cstar}(b) and \ref{fig:cstar}(c), respectively.

The results call into question the benefits of CNG at current CH$_4$ leakage estimates, given the relatively small gain of \mrsept{2\% using} a switching portfolio (from CNG to gasoline) over using gasoline alone, and the dominance of gasoline-based vehicles and infrastructure today. The investment required to make the transition to CNG may not be justified by the modest gains in energy consumption. 
Furthermore, the results demonstrate the higher energy consumption supported by the lower emissions technologies (EV, algae and switchgrass, RNG). The CNG, gasoline pair does not meet projected energy demand under this RF target.

\begin{figure*}[htbp]
\centering
\includegraphics[width=\textwidth]{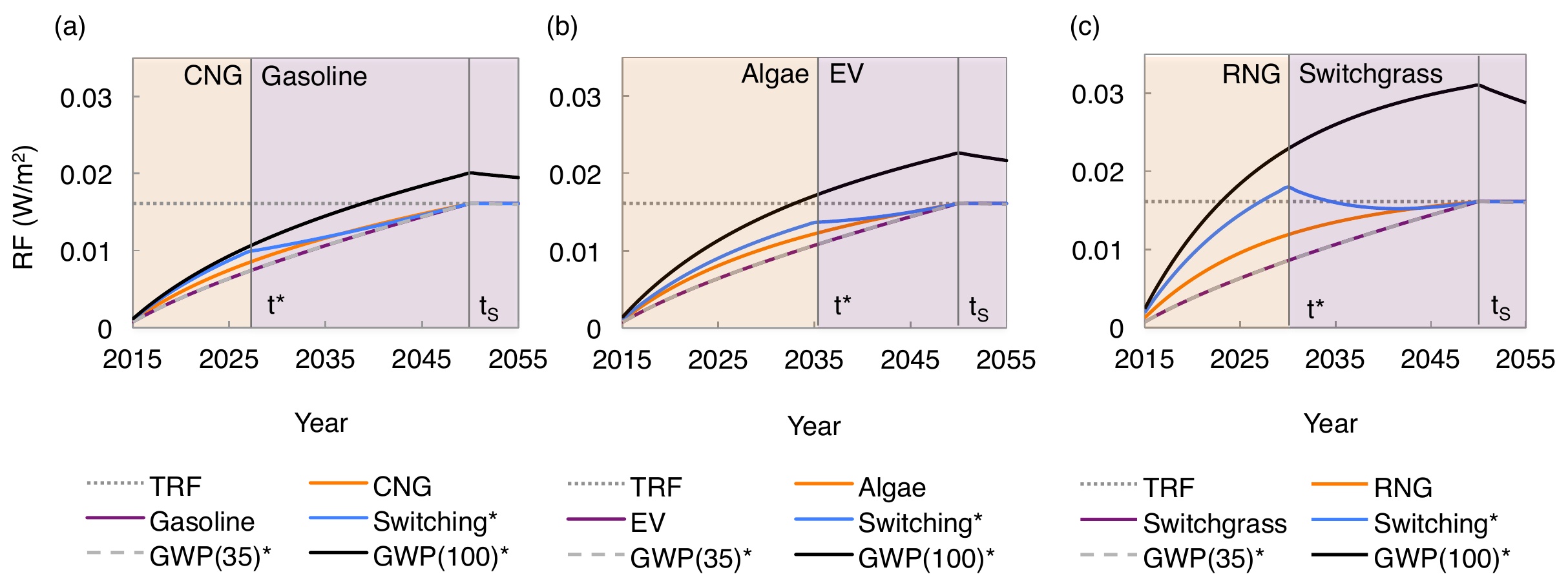}
\caption{\small{\mren{RF} resulting from technology portfolios using marginal \mren{RF} and the GWP(100) to evaluate impacts. (a) Gasoline and CNG portfolios. (b) Electric vehicle (EV) and algae portfolios. (c) Switchgrass and landfill renewable natural gas (RNG) portfolios. Switching portfolios (blue) outperform portfolios relying on CH$_4$-light (purple) or CH$_4$-heavy (orange) fuels alone, in that they allow greater energy consumption while avoiding a significant overshoot of the sector \mren{RF} target (TRF). Optimal technology choice using the GWP(100) to select the technology and energy consumption level (\roy{black}) exhibits a target overshoot over a wide range of years. \mroy{Using the GWP(35) to select the optimal technology portfolio results in the selection of the CH$_4$-light technologies (gasoline, 
EV and switchgrass) and leads to no RF target overshoots (dashed black) and less energy consumption than the corresponding switching portfolio. 
} ($t^*$, switching year; $t_S$, stabilization year).}}
\label{fig:rfstar}
\end{figure*} 

The RF trajectories for individual technologies in each pair, and their optimal switching portfolios, are shown in Fig.\ \ref{fig:rfstar}(a)-(c).
While the RF constraint is met in the stabilization year, the RNG-switchgrass optimal switching portfolio exhibits limited overshoots prior to the stabilization year, peaking at approximately 11$\%$ above the sector RF target (0.1$\%$ of the global target) in the switching year.  Early overshoots can be concerning if they are relatively large and long-lasting, and if policy targets are set to be consistent with climate system thresholds above which abrupt climate changes may occur \cite{Alley2003,Lenton2008}. 


\mrsept{An early RF overshoot can occur with a switching portfolio if the CH$_4$-heavy technology has a sufficiently high CH$_4$ intensity
relative to the CH$_4$-light technology (if the percent difference is significantly large). Early consumption using a CH$_4$-heavy technology can in this case lead to an increase in RF followed by a rapid decrease after switching, resulting in a switching year `peak' RF that exceeds the RF constraint.} 
\roysept{This effect is observed for the RNG-switchgrass portfolio, whereas the CNG-gasoline and algae-EV portfolios do not result in an RF overshoot (Fig.\ \ref{fig:rfstar}). See Sec.\ S6.}

Overshoots can be avoided by choosing an earlier stabilization year (if averse to the risk of a temporary overshoot) (Sec.\ S6.4). 
Overshoots could also be \roysept{restricted in a policy context where gases are capped separately. A `multi-basket' policy could be formulated to achieve an overshoot restriction.} In this case planning to transition from a CH$_4$-heavy to the CH$_4$-light technology still provides an advantage over applying the GWP, as the switching portfolio will allow greater energy consumption while meeting the RF constraint (see Fig.\ \ref{fig:cstar}(a) and Sec.\ S\mren{8}). 

\subsection{Optimal portfolio based on the GWP} \label{gwpresults}

The GWP-based optimal technology portfolio is determined by using the intended RF based on GWP$(\tau)$ from \eqref{rfp'} instead \roy{of} \eqref{RFpS} in the technology portfolio optimization model (see Sec.\ \roy{S7} for the general solution and proof). Using the GWP(100), the CH$_4$-heavy technologies in each pair are used to satisfy the entire portfolio, since they have lower GWP(100)-evaluated impacts than their CH$_4$-light counterparts (Fig.\ \ref{fig:rfandgwp}(b) at $\tau=100$).  The GWP(100)-based intended RF underestimates actual RF in the stabilization year, thus allowing higher energy consumption and overshooting the RF target (black lines in Fig. \ref{fig:rfstar}). If instead the RF constraint is forced to be met, the allowed energy consumption can be much lower when applying the GWP for technology evaluation than when planning for a switching portfolio (20\% lower 
using the GWP(100) with a cap versus the RNG-switchgrass portfolio, Fig.\ \ref{fig:cstar} and Sec.\ S8).

\mroy{Since the CH$_4$-light technology has a lower GWP(35) than its CH$_4$-heavy counterpart for each technology pair (Fig. \ref{fig:rfandgwp}(b) at $\tau=35$), the CH$_4$-light technology is selected over the entire horizon when the GWP(35) is applied. Because the integration horizon is the same as the stabilization horizon, the GWP(35)-based intended RF is consistent with the actual RF.  Therefore, the maximum energy consumption allowed by the GWP(35) (using Eq.\ \eqref{rfp'}) is the same as that allowed by the CH$_4$-light technologies (Eq.\ \eqref{RFpS}) while meeting the RF target (Fig. \ref{fig:cstar}, dashed grey lines in Fig. \ref{fig:rfstar}). The switching portfolio can support greater energy consumption than the GWP(35)-based} 
\mroysept{selection of the CH$_4$-light technology alone}. The RNG-switchgrass portfolio allows a 15\% energy gain over switchgrass alone.  The algae-EV portfolio allows a 12\% gain over EV alone, and CNG-gasoline allows for a 2\% gain over gasoline alone. See Fig.\ \ref{fig:cstar}(b) and Sec.\ S7. 

\mrsept{The energy gains of the switching portfolio over using the CH$_4$-light technology alone are determined by the percent difference in CH$_4$ intensities of the CH$_4$-heavy and CH$_4$-light technologies. The percent energy gains of the dynamic evaluation method can be larger if this difference is greater (as long as the difference is not so large that the CH$_4$-heavy is no longer selected over the CH$_4$-light technology). 
See Fig.\ \ref{fig:cstar} and Sec.\ S7.} 

\subsection{\jt{Robustness of results}}
\label{results2}

\jt{Here we discuss the robustness of the results presented to uncertainties in the stabilization year, the radiative efficiencies and lifetimes of greenhouse gases, and the sectoral RF target. We find that the benefits of technology transition portfolios are relatively robust to these uncertainties, suggesting the utility of this approach to technology evaluation (given a 3 W/m$^2$ global RF stabilization target) despite inherent lack of knowledge about the future. We also discuss how the insights from this research apply given changing emissions intensities of technologies, variable technology costs and quality of service, and alternative global RF stabilization targets.}  

\paragraph{Sensitivity to stabilization year uncertainty.}
\jt{We compare the optimal decisions based on a \roysept{plausible range of stabilization horizons} for a 3 W/m$^2$ RF target (Sec.\ S1). 
Examining the stochastic case, where technologies are evaluated based on the expected stabilization year (2043) but actual stabilization may occur earlier or later in the range 2035-2050, the switching portfolio still outperforms other portfolios (CH$_4$-light/GWP(35), CH$_4$-heavy, capped GWP(100), with the energy consumption gains of the switching portfolio only modestly reduced.
\mrsept{Stabilization year uncertainty can reduce the energy consumption gains (in gasoline-equivalent km) of the switching portfolios by 4-6\%. See Tab.\ S9.}   
Therefore, for these technology pairs, the performance of an optimal portfolio is relatively insensitive to uncertainty in the stabilization year. 
}  

\paragraph{Sensitivity to a changing background concentration of greenhouse gases.} 
\jt{We test the robustness of the gains of the dynamic emissions evaluation model over the static GWP, given that the radiative efficiencies \mrsept{of CO$_2$, CH$_4$, and N$_2$O} and the lifetime of CH$_4$ are likely to vary over time as the background concentrations of greenhouse gases change. To test the sensitivity of our results to these changes we select the RCP2.6 
\cite{vanVuuren2011b} as a sample scenario. 
(The RCP2.6 is just one possible 3 W/m$^2$--compliant scenario but is reasonable for demonstrating the rough scale of the effect of changing radiative efficiencies and CH$_4$ lifetimes.)} 

\jt{We find that the results are robust to \mrsept{changing radiative efficiencies and a changing CH$_4$ lifetime.} If technology switching decisions are made using the assumptions of constant radiative efficiency and CH$_4$ lifetime (representing the forward-looking part of the model), but energy consumption and RF scenarios are determined based on a changing radiative efficiency and a variable lifetime (representing the realized outcome), the gains of the switching portfolios \mrsept{are preserved.  Specifically, the gains over the CH$_4$-heavy (CH$_4$-light) technology are 17\% (4\%) for CNG-gasoine, 15\% (16\%) for algae-EV and 56\% (25\%) for RNG-switchgrass. This is compared to gains of 17\% (2\%), 12\% (12\%), and 51\% (15\%), respectively, under the constant radiative efficiency case.}  
See Sec. S10, Figs. S11 and S12.} 
\mrsept{We further examine the impact of other scenarios where greenhouse gas emissions and concentrations continue to increase over time (namely, RCP6 and RCP8.5) and find that the gains of the switching portfolio over the CH$_4$-light and CH$_4$-heavy technologies are comparable to the results shown for RCP2.6.  See Sec.\ S10.} 

\paragraph{Sensitivity to sectoral RF target.} 
\jt{Variations in the fraction of the global RF target allocated to a \mre{given} sector would change the energy consumption levels in our numerical analysis, but would not affect the technology transition (or CH$_4$ mitigation)  timeline in the optimal portfolio.  (The change in energy consumption due to a change in the RF target can be inferred from Eq.\ (S11) in Prop.\ S2. The optimal switching timeline is unaffected as Eq.\ (S6) in Prop.\ S1 is independent of the RF target. See Sec.  S6.2 and S6.1, respectively.)}

\paragraph{Effect of variable emissions intensities.} 
\jt{CH$_4$ intensities depend on venting and leakage in the production or supply infrastructure and are expected to vary across geographical locations and over time. Significant reductions may be possible \cite{EPA2013}. If the CH$_4$-emissions intensity of the `CH$_4$-heavy' technologies decreases, as compared to the current U.S. estimates on which the results are based
, switching will move closer to the intended stabilization year up to a point where switching is no longer optimal. We can estimate how significant these CH$_4$-emissions intensity reductions would need to be to make switching suboptimal. Holding other emissions intensities constant, if the CH$_4$ emissions intensity of CNG decreases by 73\% or greater, switching to gasoline is no longer optimal. Reductions of the algae (RNG) CH$_4$ emissions intensity of 46\% (66\%) would make switching to EV (switchgrass) suboptimal. See Sec.\ S6.5.}


\paragraph{Effect of technology costs or quality of service.}
\jt{The model is constructed to determine whether a dynamic emissions impact evaluation can yield significant gains, not to represent an expected outcome scenario in the real marketplace.
The numerical results we present would only hold in the marketplace \jt{(with the lowest emissions technology pair representing the optimal portfolio)} if 
technology costs and service level were comparable, energy consumption were equated to economic benefit, and the technologies examined represented the full range of options available.
Although these conditions are not met, 
the model demonstrates the potential for substantial energy consumption gains (and associated economic benefits) from a dynamic emissions impact evaluation for planning technology switching or CH$_4$ mitigation, \roysept{under a single- or multi-basket emissions cap}.
 The dynamic emissions impacts we quantify are one important input into technology decisions in the marketplace under a climate policy, but the transition timelines and choice between technology pairs would also depend on the relative costs and service limitations of technologies (e.g. EV range constraints), as well as the benefits of the energy} 
\mrsept{services provided.}


\paragraph{Effect of alternative global RF stabilization targets.}

\jt{For stabilization horizons stretching, for example, to 2100 either due to changes in assumptions regarding the range of plausible emissions reduction rates or higher RF stabilization levels, the results presented here would change, with CH$_4$-heavy technologies favored for a longer period of time. In this case, the instantaneous measure of RF impact in the stabilization year could result in substantial RF overshoots in earlier years, making the emissions and technology evaluation approach described here less attractive. 
We note that century scale horizons are longer than practical for technology planning, and are therefore outside the scope of this paper. Furthermore, planning for an earlier-than-realized stabilization year would reduce the risk of radiative forcing overshoots.} 

\section{Conclusion and Discussion}\label{conclusion}

In this paper we focus on dynamic technology evaluation and choice to meet a given \mren{RF} stabilization level. We show that the optimal choice can be a technology switching portfolio, where the CH$_4$-heavy technology is used initially, followed by a switch to a CH$_4$-light option. Such a switching portfolio can allow greater energy consumption than the exclusive use of either technology.  These results support the case for using appropriate CH$_4$-heavy bridging technologies, given a sufficiently long stabilization horizon, but also caution against using CH$_4$-heavy technologies too close to the stabilization time-frame. We note that the same benefits would apply to planning for reducing CH$_4$ leakage from technologies that are currently CH$_4$-heavy but show potential for decreasing CH$_4$ emissions \cite{Levi2013,Brandt2014}. 
This result points to two options: transitioning to low-CH$_4$ technologies or mitigating CH$_4$ emissions. 

The model demonstrates the benefits, as compared to the static GWP($\tau$), of planning for technology transitions using a dynamic emissions impact evaluation approach or the ICI metric \cite{Edwards2014}. 
\mrsept{A switching portfolio can allow greater energy consumption than the GWP(35)-selected, CH$_4$-light technology (up to 15\% for the technology examples studied),} while still meeting the RF 
target. 
The GWP(100)-based portfolio \mrsept{allows even greater energy consumption but can} lead to a significant overshoot of the intended RF stabilization level. 
If an RF constraint is applied exogenously, \roysept{a situation which approximates the real-world case of a multi-basket emissions policy that regulates different gases through separate caps,} the \roy{GWP(100)-based selection 
allows less energy consumption than the technology switching portfolio} (up to 20\% less for the technology examples studied). 
 
The technology evaluation approach we develop is designed to be robust to uncertainty regarding the stabilization scenario but does require specification of a global \mren{RF} stabilization target. Given a stabilization level, a range of 
 stabilization years is determined, and the optimal year for switching from a CH$_4$-heavy to a CH$_4$-light technology is well-defined by this range. The optimal switching year does not depend on the future energy consumption level, and the benefits of technology switching will apply across a wide range of possible energy consumption scenarios. 
This robustness is important because there is a critical need for technology evaluation tools that perform well despite inherent uncertainty about the future, in order to inform technology design, private investment decisions, and policy development. These tools should also be transparent and easy to use and yet, to perform well, should incorporate broader climate policy goals. 

We present such an approach here, to inform technology development timelines. Such planning can help direct efforts to reduce the costs of low-emissions intensity technologies \cite{Trancik2013,Bettencourt2013,Farmer07b,Nagy12}. 
\jt{Private actors investing in R\&D and technology production capacity might use the insights on CH$_4$ mitigation timelines to decide which technology designs to invest in. Public actors might use the results to evaluate projects for R\&D funding, considering likely technology-to-market development timelines and CH$_4$ leakage rates. \roysept{The results also point to the importance of incorporating dynamic emissions impacts or multi-basket emissions caps into emissions regulations, to avoid the potential RF overshoots resulting from applying the GWP(100).} US EPA regulations on power plants \cite{EPA2014a,EPA2014b}, and other current and proposed policies \cite{EOP2014,Whitehouse2013,Trancik2014}, rely on the GWP(100) to evaluate technology impacts -- or do not account for the impacts of non-CO$_2$ greenhouse gases at all. 
Methods like the one we propose can inform the formulation of policies to meet the demand for energy while also meeting climate change mitigation goals.}

\section{Acknowledgements}\label{acknowledge}

We thank the MIT Energy Initiative for supporting this research through the MITEI Seed Fund Program. We also thank the Reed Foundation for supporting this work. MRE acknowledges support from a National Science Foundation Graduate Research Fellowship under Grant No. 1122374. We thank three anonymous reviewers for their helpful suggestions.


\normalem
\bibliographystyle{unsrtnat}
\bibliography{portbib,portbibnew}

\begin{thebibliography}{56}
\providecommand{\natexlab}[1]{#1}
\providecommand{\url}[1]{\texttt{#1}}
\expandafter\ifx\csname urlstyle\endcsname\relax
  \providecommand{\doi}[1]{doi: #1}\else
  \providecommand{\doi}{doi: \begingroup \urlstyle{rm}\Url}\fi

\bibitem[{Stocker, T. F. and Qin, D. and Plattner, G. and Tignor, M. M. B. and
  Allen, S. K. and Boschung, J. and Nauels, A. and Xia, Y. and Bex, V. and
  Midgley, P. M.}(2013)]{IPCC2013}
{Stocker, T. F. and Qin, D. and Plattner, G. and Tignor, M. M. B. and Allen, S.
  K. and Boschung, J. and Nauels, A. and Xia, Y. and Bex, V. and Midgley, P.
  M.}, editor.
\newblock \emph{{Climate Change 2013: The Physical Science Basis}}.
\newblock Cambridge University Press, 2013.

\bibitem[UNF(1992)]{UNFCCC1992}
\emph{{United Nations Framework Convention on Climate Change}}.
\newblock May 1992.
\newblock S. Treaty Doc No. 102-38, 1771 U.N.T.S. 107.

\bibitem[{United Nations Framework Convention on Climate
  Change}(2010)]{UNFCCC2009}
{United Nations Framework Convention on Climate Change}.
\newblock \emph{{Report of the Conference of the Parties on its fifteenth
  session, held in Copenhagen from 7 to 19 December 2009}}.
\newblock March 2010.
\newblock {FCCC/CP/2009/11/Add.1}.

\bibitem[Jackson(2009)]{Jackson2009}
S.~Jackson.
\newblock Parallel pursuit of near-term and long-term climate mitigation.
\newblock \emph{Science}, 326:\penalty0 526--527, 2009.

\bibitem[Smith et~al.(2012)Smith, Lowe, Bowerman, Gohar, and Allen]{Smith2012}
S.~M. Smith, J.~A. Lowe, N.~H.~A. Bowerman, L.~K. Gohar, and M.~R. Allen.
\newblock Equivalence of greenhouse-gas emissions for peak temperature limits.
\newblock \emph{Nature Clim. Change}, 2:\penalty0 535--538, 2012.

\bibitem[Shindell et~al.(2012)Shindell, Kuylenstierna, Vignati, Dingenen,
  Amann, Klimont, Anenberg, Muller, Janssens-maenhout, Raes, Schwartz,
  Faluvegi, Pozzoli, Kupiainen, Hoglund-isaksson, Emberson, Streets,
  Ramanathan, Hicks, Oanh, Milly, Williams, Demkine, and Fowler]{Shindell2012}
D.~Shindell, J.~C.~I. Kuylenstierna, E.~Vignati, R.~V. Dingenen, M.~Amann,
  Z.~Klimont, S.~C. Anenberg, N.~Muller, G.~Janssens-maenhout, F.~Raes,
  J.~Schwartz, G.~Faluvegi, L.~Pozzoli, K.~Kupiainen, L.~Hoglund-isaksson,
  L.~Emberson, D.~Streets, V.~Ramanathan, K.~Hicks, N.~T.~K. Oanh, G.~Milly,
  M.~Williams, V.~Demkine, and D.~Fowler.
\newblock Simultaneously mitigating near-term climate change and improving
  human health and food security.
\newblock \emph{Science}, 335:\penalty0 183--189, 2012.

\bibitem[Shoemaker et~al.(2013)Shoemaker, Schrag, Molina, and
  Ramanathan]{Shoemaker2013b}
J.~K. Shoemaker, D.~P. Schrag, M.~J. Molina, and V.~Ramanathan.
\newblock What role for short-lived climate pollutants in mitigation policy?
\newblock \emph{Science}, 342:\penalty0 1323--1324, 2013.

\bibitem[Rodhe(1990)]{Rodhe1990}
H.~Rodhe.
\newblock A comparison of the contribution of various gases to the greenhouse
  effect.
\newblock \emph{Science}, 248\penalty0 (4960):\penalty0 1217--1219, 1990.

\bibitem[Shindell et~al.(2009)Shindell, Faluvegi, Koch, Schmidt, Unger, and
  Bauer]{Shindell2009}
Drew~T. Shindell, Greg Faluvegi, Dorothy~M. Koch, Gavin~A. Schmidt, Nadine
  Unger, and Susanne~E. Bauer.
\newblock Improved attribution of climate forcing to emissions.
\newblock \emph{Science}, 326\penalty0 (5953):\penalty0 716--718, 2009.

\bibitem[{O'Neill}(2000)]{ONeill2000}
B.~C. {O'Neill}.
\newblock The jury is still out on global warming potentials.
\newblock \emph{Climatic Change}, 44:\penalty0 427--443, 2000.

\bibitem[O'Neill(2003)]{ONeill2003}
B.~C. O'Neill.
\newblock Economics, natural science, and the costs of global warming
  potentials.
\newblock \emph{Climatic Change}, 58:\penalty0 251--260, 2003.

\bibitem[Shine(2009)]{Shine2009}
Keith~P. Shine.
\newblock The global warming potential -- the need for an interdisciplinary
  retrial.
\newblock \emph{Climatic Change}, 96\penalty0 (4):\penalty0 467--472, 2009.

\bibitem[Daniel et~al.(2011)Daniel, Solomon, Sanford, McFarland, Fuglestvedt,
  and Friedlingstein]{Daniel2011}
J.~S. Daniel, S.~Solomon, T.~J. Sanford, M.~McFarland, J.~S. Fuglestvedt, and
  P.~Friedlingstein.
\newblock Limitations of single-basket trading: Lessons from the {Montreal
  Protocol} for climate policy.
\newblock \emph{Climatic Change}, 111\penalty0 (2):\penalty0 241--248, 2011.

\bibitem[Peters et~al.(2011)Peters, Aamaas, Lund, Solli, and
  Fuglestvedt]{Peters2011}
G.~P. Peters, B.~Aamaas, M.~T. Lund, C.~Solli, and J.~S. Fuglestvedt.
\newblock Alternative ``global warming'' metrics in life cycle assessment: A
  case study with existing transportation data.
\newblock \emph{Environ.\:Sci.\:Technol.}, 45\penalty0 (20):\penalty0
  8633--8641, 2011.

\bibitem[Stratton et~al.(2011)Stratton, Wolfe, and Hileman]{Stratton2011}
R.~W. Stratton, P.~J. Wolfe, and J.~I. Hileman.
\newblock Impact of aviation non-{CO}$_2$ combustion effects on the
  environmental feasibility of alternative jet fuels.
\newblock \emph{Environ. Sci. Technol.}, 45\penalty0 (24):\penalty0
  10736--10743, 2011.

\bibitem[Kendall(2012)]{Kendall2012}
A.~Kendall.
\newblock Time-adjusted global warming potentials for {LCA} and carbon
  footprints.
\newblock \emph{Int. J. Life Cycle Ass.}, 17\penalty0 (8):\penalty0 1042--1049,
  2012.

\bibitem[Edenhofer et~al.(2012)Edenhofer, Pichs~Madruga, and Sokona]{IPCC2012}
O.~Edenhofer, R.~Pichs~Madruga, and Y.~Sokona, editors.
\newblock \emph{Renewable Energy Sources and Climate Change Mitigation: Special
  Report of the Intergovernmental Panel on Climate Change}.
\newblock {Cambridge University Press}, 2012.

\bibitem[Whi(2013)]{Whitehouse2013}
\emph{The President's Climate Action Plan}.
\newblock Executive Office of the President, 2013.

\bibitem[{U.S. Environmental Protection Agency}(2014{\natexlab{a}})]{EPA2014a}
{U.S. Environmental Protection Agency}.
\newblock Standards of performance for greenhouse gas emissions from new
  stationary sources: Electric utility generating units.
\newblock \emph{Federal Register}, 79\penalty0 (5), 2014{\natexlab{a}}.

\bibitem[{U.S. Environmental Protection Agency}(2014{\natexlab{b}})]{EPA2014b}
{U.S. Environmental Protection Agency}.
\newblock Carbon pollution emission guidelines for existing stationary sources:
  Electric utility generation units.
\newblock \emph{Federal Register}, 79\penalty0 (117), 2014{\natexlab{b}}.

\bibitem[of~the President(2014)]{EOP2014}
Executive~Office of~the President.
\newblock \emph{Strategy to Reduce Methane Emissions}.
\newblock March 2014.

\bibitem[van Vuuren et~al.(2006)van Vuuren, Weyant, and de~la
  Chesnaye]{vanVuuren2006}
D.~P. van Vuuren, J.~Weyant, and F.~de~la Chesnaye.
\newblock Multi-gas scenarios to stabilize radiative forcing.
\newblock \emph{Energy Econ.}, 28:\penalty0 102--120, 2006.

\bibitem[Weyant et~al.(2006)Weyant, de~la Chesnaye, and Blanford]{Weyant2006}
J.~P. Weyant, F.~C. de~la Chesnaye, and G.~J. Blanford.
\newblock Overview of {EMF-21}: Multigas mitigation and climate policy.
\newblock \emph{Energy J.}, 27:\penalty0 1--32, 2006.

\bibitem[van Vuuren et~al.(2007)van Vuuren, den Elzen, Lucas, Eickhout,
  Strengers, van Ruijven, Wonink, and van Houdt]{vanVuuren2007}
D.~P. van Vuuren, M.~G.~J. den Elzen, P.~L. Lucas, B.~Eickhout, B.~J.
  Strengers, B.~van Ruijven, S.~Wonink, and R.~van Houdt.
\newblock Stabilizing greenhouse gas concentrations at low levels: An
  assessment of reduction strategies and costs.
\newblock \emph{Climatic Change}, 81:\penalty0 119---159, 2007.

\bibitem[Rogelj et~al.(2011)Rogelj, Hare, Lowe, van Vuuren, Riahi, Matthews,
  Hanaoka, Jiang, and Meinshausen]{Rogelj2011}
J.~Rogelj, W.~Hare, J.~Lowe, D.~P. van Vuuren, K.~Riahi, B.~Matthews,
  T.~Hanaoka, K.~Jiang, and M.~Meinshausen.
\newblock Emissions pathways consistent with a 2$^{\circ}${C} global
  temperature limit.
\newblock \emph{Nature Clim. Change}, 1:\penalty0 413--418, 2011.

\bibitem[Smith et~al.(2013)Smith, Karas, Edmonds, Eom, and
  Mizrahi]{Smith2013_1}
S.~J. Smith, J.~Karas, J.~Edmonds, J.~Eom, and A.~Mizrahi.
\newblock Sensitivity of multi-gas climate policy to emission metrics.
\newblock \emph{Climatic Change}, 117:\penalty0 663--675, 2013.

\bibitem[Rogelj et~al.(2014)Rogelj, Schaeffer, Meinshausen, Shindell, Hare,
  Klimont, Velders, Amann, and Schellnhuber]{Rogelj2014}
J.~Rogelj, M.~Schaeffer, M.~Meinshausen, D.~T. Shindell, W.~Hare, Z.~Klimont,
  G.~J.~M. Velders, M.~Amann, and H.~J. Schellnhuber.
\newblock Disentangling the effects of {CO}$_2$ and short-lived climate forcer
  mitigation.
\newblock \emph{Proc. Natl. Acad. Sci. USA}, 111\penalty0 (46):\penalty0
  16325--16330, 2014.

\bibitem[Smith and Mizrahi(2013)]{Smith2013_2}
S.~J. Smith and A.~Mizrahi.
\newblock Near-term mitigation by short-lived forcers.
\newblock \emph{Proc. Natl. Acad. Sci. USA}, 110\penalty0 (35):\penalty0
  14202--14206, 2013.

\bibitem[Bowerman et~al.(2013)Bowerman, Frame, Huntingford, Lowe, Smith, and
  Allen]{Bowerman2013}
N.~H.~A. Bowerman, D.~J. Frame, C.~Huntingford, J.~A. Lowe, S.~M. Smith, and
  M.~R. Allen.
\newblock The role of short-lived climate pollutants in meeting temperature
  goals.
\newblock \emph{Nature Clim. Change}, 3\penalty0 (12):\penalty0 1021--1024,
  2013.

\bibitem[Shoemaker and Schrag(2013)]{Shoemaker2013}
J.~K. Shoemaker and D.~P. Schrag.
\newblock The danger of overvaluing methane's influence on future climate
  change.
\newblock \emph{Climatic Change}, 120\penalty0 (4):\penalty0 903--914, 2013.

\bibitem[Trancik and Cross-Call(2013)]{Trancik2013}
J.~E. Trancik and D.~Cross-Call.
\newblock Energy technologies evaluated against climate targets using a cost
  and carbon trade-off curve.
\newblock \emph{Environ.\:Sci.\:Technol.}, 47\penalty0 (12):\penalty0
  6673--6680, 2013.

\bibitem[Trancik et~al.(2014)Trancik, Chang, Karapataki, and
  Stokes]{Trancik2014}
J.~E. Trancik, M.~T. Chang, C.~Karapataki, and L.~C. Stokes.
\newblock Effectiveness of a segmental approach to climate policy.
\newblock \emph{Environ.\:Sci.\:Technol.}, 48\penalty0 (1):\penalty0 27--35,
  2014.

\bibitem[Manne and Richels(2001)]{Manne2001}
A.~S. Manne and R.~G. Richels.
\newblock An alternative approach to establishing trade-offs among greenhouse
  gases.
\newblock \emph{Nature}, 410\penalty0 (6829):\penalty0 675--677, 2001.

\bibitem[Shine et~al.(2007)Shine, Berntsen, Fuglestvedt, Skeie, and
  Stuber]{Shine2007}
K.~P. Shine, T.~K. Berntsen, J.~S. Fuglestvedt, R.~B. Skeie, and N.~Stuber.
\newblock Comparing the climate effect of emissions of short- and long-lived
  climate agents.
\newblock \emph{Phil. Trans. R. Soc. A}, 365\penalty0 (1856):\penalty0
  1903--1914, 2007.

\bibitem[{Plattner, G.-K. and Stocker, T. and Midgley, P. and Tignor,
  M.}(2009)]{IPCC2009}
{Plattner, G.-K. and Stocker, T. and Midgley, P. and Tignor, M.}, editor.
\newblock \emph{{IPCC Expert Meeting on the Science of Alternative Metrics}}.
\newblock {Intergovernmental Panel on Climate Change}, 2009.

\bibitem[Johansson(2012)]{Johansson2012}
D.~J.~A. Johansson.
\newblock Economics- and physical-based metrics for comparing greenhouse gases.
\newblock \emph{Climatic Change}, 110\penalty0 (1-2):\penalty0 123--141, 2012.

\bibitem[Tanaka et~al.(2013)Tanaka, Johansson, O'Neill, and
  Fuglestvedt]{Tanaka2013}
K.~Tanaka, D.~J.~A. Johansson, B.~C. O'Neill, and J.~S. Fuglestvedt.
\newblock Emissions metrics under a 2$^{\circ}${C} stabilization target.
\newblock \emph{Climatic Change}, 117\penalty0 (4):\penalty0 933--941, 2013.

\bibitem[Berntsen et~al.(2010)Berntsen, Tanaka, and Fuglestvedt]{Berntsen2010}
T.~Berntsen, K.~Tanaka, and J.~S. Fuglestvedt.
\newblock Does black carbon abatement hamper {CO}$_2$ abatement?
\newblock \emph{Climatic Change}, 103\penalty0 (3):\penalty0 627--633, 2010.

\bibitem[Alvarez et~al.(2012)Alvarez, Pacala, Winebrake, Chameides, and
  Hamburg]{Alvarez2012}
R.~A. Alvarez, S.~W. Pacala, J.~J. Winebrake, W.~L. Chameides, and S.~P.
  Hamburg.
\newblock Greater focus needed on methane leakage from natural gas
  infrastructure.
\newblock \emph{Proc. Natl Acad. Sci. USA}, 109\penalty0 (17):\penalty0
  6435--6440, 2012.

\bibitem[Edwards and Trancik(2014)]{Edwards2014}
M.~R. Edwards and J.~E. Trancik.
\newblock Climate impacts of energy technologies depend on emissions timing.
\newblock \emph{Nature Clim. Change}, 4:\penalty0 347--352, 2014.

\bibitem[Lenton(2011)]{Lenton2011}
T.~M. Lenton.
\newblock Beyond 2$^{\circ}${C}: {Redefining} dangerous climate change for
  physical systems.
\newblock \emph{WIREs Clim.\:Change}, 2\penalty0 (3):\penalty0 451--461, 2011.

\bibitem[Allen et~al.(2009)Allen, Frame, Huntingford, Jones, Lowe, Meinshausen,
  and Meinshausen]{Allen2009}
M.~R. Allen, D.~J. Frame, C.~Huntingford, C.~D. Jones, J.~A. Lowe,
  M.~Meinshausen, and N.~Meinshausen.
\newblock Warming caused by cumulative carbon emissions towards the trillionth
  tonne.
\newblock \emph{Nature}, 458\penalty0 (7242):\penalty0 1163--1166, 2009.

\bibitem[Meinhausen et~al.(2011)Meinhausen, Smith, Calvin, Daniel, Kainuma,
  Lamarque, Matsumoto, Montzka, Raper, Riahi, Thonson, Velders, and van
  Vuuren]{Meinhausen2011}
M.~Meinhausen, S.~J. Smith, K.~Calvin, J.~S. Daniel, M.~L.~T. Kainuma, J.~F.
  Lamarque, K.~Matsumoto, S.~A. Montzka, S.~C.~B. Raper, K.~Riahi, A.~Thonson,
  G.~J.~M. Velders, and D.~P.~P. van Vuuren.
\newblock The {RCP} greenhouse gas concentrations and their extensions from
  1765 to 2300.
\newblock \emph{Climatic Change}, 109:\penalty0 213--241, 2011.

\bibitem[van Vuuren et~al.(2011{\natexlab{a}})van Vuuren, J., Kainuma, Riahi,
  Thomson, Hibbard, Hurtt, Kram, Krey, Lamarque, Masui, Meinhausen,
  Nakicenovic, Smith, and Rose]{vanVuuren2011b}
D.~P. van Vuuren, Edmonds J., M.~Kainuma, K.~Riahi, A.~Thomson, K.~Hibbard,
  G.~C. Hurtt, T.~Kram, V.~Krey, J.~Lamarque, T.~Masui, M.~Meinhausen,
  N.~Nakicenovic, S.~J. Smith, and S.~K. Rose.
\newblock {The representative concentration pathways: and overview}.
\newblock \emph{Climatic Change}, 109:\penalty0 5---31, 2011{\natexlab{a}}.

\bibitem[{U.S. Energy Information Administration}(2014)]{EIA2014}
{U.S. Energy Information Administration}.
\newblock \emph{{Annual Energy Outlook 2014}}.
\newblock April 2014.

\bibitem[{Myhre, G. and Shindell, D., editor}(2013)]{IPCC2013Ch8}
{Myhre, G. and Shindell, D., editor}.
\newblock {Anthropogenic and Natural Radiative Forcing}.
\newblock In \emph{Climate Change 2013: The Physical Science Basis.
  Contribution of Working Group I to the Fifth Assessment Report of the
  Intergovernmental Panel on Climate Change}. Cambridge University Press, 2013.

\bibitem[{U.S. Environmental Protection Agency}(2014{\natexlab{c}})]{EPA2013}
{U.S. Environmental Protection Agency}.
\newblock {Global Mitigation of Non-CO$_2$ Greenhouse Gases: 2010-2030}.
\newblock \emph{Federal Register}, 79\penalty0 (5), 2014{\natexlab{c}}.

\bibitem[van Vuuren et~al.(2011{\natexlab{b}})van Vuuren, Stehfest, den Elzen,
  Kram, van Vliet, Deetman, Isaac, Goldewijke, Hof, Beltran, Oostenrijk, and
  van Ruijven]{vanVuuren2011a}
D.~van Vuuren, E.~Stehfest, M.~G.~J. den Elzen, T.~Kram, J.~van Vliet,
  S.~Deetman, M.~Isaac, K.~K. Goldewijke, A.~Hof, A.~M. Beltran, R.~Oostenrijk,
  and B.~van Ruijven.
\newblock {RCP}2.6: Exploring the possibility to keep global mean temperature
  increase below 2$^{\circ}${C}.
\newblock \emph{Climatic Change}, 109:\penalty0 95--116, 2011{\natexlab{b}}.

\bibitem[Williams et~al.(2012)Williams, DeBenedictis, Ghanada, Mahone, Moore,
  Morrow, Price, and Torn]{Williams2012}
J.~H. Williams, A.~DeBenedictis, R.~Ghanada, A.~Mahone, J.~Moore, W.~R. Morrow,
  S.~Price, and M.~S. Torn.
\newblock The technology path to deep greenhouse gas emissions cuts by 2050.
\newblock \emph{Science}, 335\penalty0 (6064):\penalty0 53--59, 2012.

\bibitem[Alley et~al.(2003)Alley, Marotzke, Nordhaus, Overpeck, Peteet, Pielke,
  Pierrehumbert, Rhines, Stocker, Talley, and Wallace]{Alley2003}
R.~B. Alley, J.~Marotzke, W.~D. Nordhaus, J.~T. Overpeck, D.~M. Peteet, R.~A.
  Pielke, R.~T. Pierrehumbert, P.~B. Rhines, T.~F. Stocker, L.~D. Talley, and
  J.~M. Wallace.
\newblock Abrupt climate change.
\newblock \emph{Science}, 299\penalty0 (5615):\penalty0 2005--2010, 2003.

\bibitem[Lenton et~al.(2008)Lenton, Held, Kriegler, Hall, Lucht, Rahmstorf, and
  Schellnhuber]{Lenton2008}
T.~M. Lenton, H.~Held, E.~Kriegler, J.~W. Hall, W.~Lucht, S.~Rahmstorf, and
  H.~J. Schellnhuber.
\newblock Tipping elements in the {Earth's} climate system.
\newblock \emph{Proc. Natl Acad. Sci. USA}, 105\penalty0 (6):\penalty0
  1786--1793, 2008.

\bibitem[Levi(2013)]{Levi2013}
M.~Levi.
\newblock Climate consequences of natural gas as a bridge fuel.
\newblock \emph{Climatic Change}, 118:\penalty0 609--623, 2013.

\bibitem[Brandt et~al.(2014)Brandt, Heath, Kort, O'Sullivan, Petron, Jordaan,
  Tans, Wilcox, Gopstein, Arent, Wofsy, Brown, Bradley, Stucky, and
  Harriss]{Brandt2014}
A.~R. Brandt, G.~A. Heath, E.~A. Kort, F.~O'Sullivan, G.~Petron, S.~M. Jordaan,
  P.~Tans, J.~Wilcox, A.~M. Gopstein, D.~Arent, S.~Wofsy, N.~J. Brown,
  R.~Bradley, Eardley~D. Stucky, G.~D., and R.~Harriss.
\newblock Methane leaks from north american natural gas systems.
\newblock \emph{Science}, 343:\penalty0 733--735, 2014.

\bibitem[Bettencourt et~al.(2013)Bettencourt, Trancik, and
  Kaur]{Bettencourt2013}
L.~M.~A. Bettencourt, J.~E. Trancik, and J.~Kaur.
\newblock Determinants of the pace of global innovation in energy technologies.
\newblock \emph{PLoS ONE}, 8\penalty0 (10):\penalty0 e67864, 10 2013.
\newblock \doi{10.1371/journal.pone.0067864}.

\bibitem[Farmer and Trancik(2007)]{Farmer07b}
J.~D. Farmer and J.~E. Trancik.
\newblock Dynamics of technological development in the energy sector.
\newblock In J.~P. Onstwedder and M.~Mainelli, editors, \emph{London Accord
  Final Publication}, 2007.

\bibitem[{Nagy} et~al.(2013){Nagy}, {Farmer}, {Bui}, and {Trancik}]{Nagy12}
B.~{Nagy}, J.~D. {Farmer}, Q.~M. {Bui}, and J.~E. {Trancik}.
\newblock Statistical basis for predicting technological progress.
\newblock \emph{PLoS ONE}, 8:\penalty0 e52669, 2013.

\end{thebibliography}
\end{document}